\documentclass[fleqn,twoside]{article}
\usepackage{espcrc2}
\usepackage{graphicx}

\title{Unified Approach to Dense Matter}

\author{Byung-Yoon Park\address[CNU]{CSSM, University of Adelaide,
          Adelaide 5005, Australia}%
          \thanks{On sabbatical leave of absence from Department of Physics,
                  Chungnam National University, Daejon 305-764, Korea},
        Hee-Jung Lee\address[VU]{Departament de F\'{i}sica Te\`{o}rica
          and Institut de F\'{i}sica Corpuscular
          Universitat de Val\`{e}ncia and Consejo Superior de
          Investigaciones Cient\`{i}ficas, E-46100 Burjassot
          (Val\`{e}ncia), Spain},
        Vicente Vento\addressmark,
        Joon-Il Kim\address[SNU]{Department of Physics, Seoul National University,
                Seoul 151-742, Korea},
        Dong-Pil Min\addressmark
        and
        Mannque Rho\address[HU]{Department of Physics, Hanyang University,
           Seoul 133-791, Korea, and Service de Physique Th\'{e}orique,
           CE Saclay, 91191 Gif-sur-Yvette, France}
        }
\begin{document}

\begin{abstract}
We apply the Skyrme model to dense hadronic matter, which provides a
unified approach to high density, valid in the large $N_c$ limit.
In our picture, dense hadronic matter is described by the {\em
classical} soliton configuration with minimum energy for the given
baryon number density. By incorporating the meson fluctuations on
such ground state we obtain an effective Lagrangian for meson
dynamics in a dense medium. Our starting point has been the Skyrme
model defined in terms of pions, thereafter we have extended and
improved the model by incorporating other degrees of freedom such
as dilaton, kaons and vector mesons.
\vspace{1pc}
\end{abstract}

\maketitle

\section{Introduction}

At high temperature and/or density, hadrons are expected to
possess properties that are very different from those at normal
conditions. Understanding the properties of hadrons in such
extreme conditions is currently an important issue not only in
nuclear and particle physics but also in many other related fields
such as astrophysics. Data from the high energy heavy ion
colliders, astronomical observations on compact stars and some
theoretical developments have shown that the phase diagram of
hadronic matter is far richer and more interesting than initially
expected. Lattice QCD calculations have been carried out
successfully at high temperature, however similar calculations at
high density have not yet been possible.
Theoretical developments have unveiled such interesting QCD phases
as color superconductivity. Moreover effective theories
can be derived for these extreme conditions, using macroscopic
degrees of freedom, by matching them to QCD at a scale close to
the chiral scale $\Lambda_\chi \sim 4\pi f_\pi \sim 1$
GeV.

\renewcommand{\thefootnote}{\fnsymbol{footnote}}
\setcounter{footnote}{1} We have followed a different path to
dense matter
studies\cite{PMRV02,LPMRV03,LPRV03a,LPRV03b,PRV03,KPMRV04}\footnotemark
\footnotetext{Due to the limitation in length, we only refer to
our work, on which the talk is based, and to a few others whose
results are quoted explicitly.} by using as our starting point a
model Lagrangian, in the spirit of Skyrme, which describes
hadronic matter and meson dynamics respecting the symmetries of
QCD. The parameters of the model are fixed by meson dynamics at
{\em zero} baryon number density. {\it \`A la\/} Skyrme, baryons
arise from a soliton solution, the skyrmion, with the topological
winding number describing the baryon number. In our scheme dense
matter is approximated by a system of skyrmions with a given
baryon number density whose ground state arises as a crystal
configuration\cite{SkyrmionCrystal}. Starting from this ground
state our approach provides insight on the intrinsic in-medium
dependence of meson dynamics. We have studied (i) the in-medium
properties of the mesons and (ii) the role of the other degrees of
freedom besides pions in the description of matter as it becomes
denser.

%%%%%%%%%%%%%%%%%%%%%%%%%%%%%%%%%%%%%%%%%%%%%%%%%%%%%%%%%%%%%%%%%%%%%%%
\section{Model Lagrangian}
The original Skyrme model Lagrangian\cite{Sk62} reads
\begin{equation}
\begin{array}{rcl}
{L}_{\pi} &=& \displaystyle
- \frac{f_\pi^2}{4} \mbox{Tr} (L_\mu L^\mu)
+ \frac{1}{32e^2} \mbox{Tr} [ L_\mu, L_\nu]^2 \\
&&\displaystyle
+ \frac{f_\pi^2 m_\pi^2}{4} \mbox{Tr} (U + U^\dagger - 2),
\end{array}
\label{L:p}
\end{equation}
where $L_\mu = U^\dagger \partial_\mu U$ and
$U=\exp(i\vec{\tau}\cdot\vec{\pi}) \in SU(2)$ is a nonlinear
realization of the pion fields, $f_\pi$  the decay constant and
$m_\pi$ the pion mass. The second term with $e$, the Skyrme
parameter, was introduced to stabilize the soliton solution.

The dilaton field $\chi$ can be incorporated in the model to make it
consistent with the scale anomaly of QCD\cite{EL85,BR91}. The
Lagrangian (\ref{L:p}) then becomes modified as
\begin{equation}
\begin{array}{rcl}
{L}_{\pi\chi}
&=& \displaystyle
- \frac{f_\pi^2}{4} \left(\frac{\chi}{f_\chi}\right)^2
            \mbox{Tr} (L_\mu L^\mu) \\
&& \displaystyle
+ \frac{1}{32e^2} \mbox{Tr} [ L_\mu, L_\nu]^2  \\
&& \displaystyle
+ \frac{f_\pi^2 m_\pi^2}{4} \left(\frac{\chi}{f_\chi}\right)^3
            \mbox{Tr} (U + U^\dagger - 2) \\
&&
+ \frac12 \partial_\mu \chi \partial^\mu \chi - V(\chi),
\end{array}
\label{L:pc}
\end{equation}
Note the different powers of $(\chi/f_\chi)$ in front of each
term. The last line is  the Lagrangian for the free dilaton field,
where $V(\chi)=(m_\chi^2 f_\chi^2/4)((\chi/f_\chi)^4
(\mbox{ln}(\chi/f_\chi)-\frac{1}{4})-\frac{1}{4})$, $m_\chi$ is
the dilaton mass and $f_\chi$ its decay constant.

The vector mesons, $\rho$ and $\omega$, can be included into
the model as dynamical gauge bosons of a hidden local gauge
symmetry which requires the doubling of the degrees of freedom as
$U=\xi^\dagger_L \xi_R$. One of such Lagrangian is,
for example\cite{HLS},
\begin{equation}
\begin{array}{l}
{L}_{\pi\chi\rho\omega}
= \displaystyle
\frac{f_\pi^2}{4} \left(\frac{\chi}{f_\chi}\right)^2
  \mbox{Tr}(\xi^{}_L {\cal D}_\mu \xi^\dagger_L
           -\xi^{}_R {\cal D}_\mu \xi^\dagger_R)^2 \\
\hskip 3em \displaystyle
+a\frac{f_\pi^2}{4} \left(\frac{\chi}{f_\chi}\right)^2
  \mbox{Tr}(\xi^{}_L {\cal D}_\mu \xi^\dagger_L
           +\xi^{}_R {\cal D}_\mu \xi^\dagger_R)^2 \\
\hskip 3em \displaystyle
+\frac{f_\pi^2 m_\pi^2}{4} \left(\frac{\chi}{f_\chi}\right)^3
\mbox{Tr}(\xi^\dagger_L \xi^{}_R + \xi^\dagger_R \xi_L^{} -2 ) \\
\hskip 3em \displaystyle
+ \frac{N_cg}{2} \omega_\mu B^\mu
- \frac{1}{4}
\vec{\rho}_{\mu\nu} \cdot \vec{\rho}^{\mu\nu}
-\frac{1}{4}  \omega_{\mu\nu} \omega^{\mu\nu} \\
\hskip 3em
 + \frac{1}{2} \partial_\mu \chi \partial^\mu \chi
 -V(\chi),
\end{array}
\label{L:pcrw}
\end{equation}
with ${\cal D}_\mu=\partial_\mu-\frac{i}{2}\vec{\tau}\cdot\vec{\rho}_\mu
-\frac{i}{2}\omega_\mu$, $\vec{\rho}_{\mu\nu} = \partial_\mu \vec{\rho}_\nu
 - \partial_\nu \vec{\rho}_\mu + g \vec{\rho}_\mu \times \vec{\rho}_\nu$,
$\omega_{\mu\nu}=\partial_\mu \omega_\nu -
\partial_\nu\omega_\mu$, and where $B_\mu$ is the {\em topological}
baryon number current. The quartic Skyrme term of (\ref{L:p}) is
not present, because its stabilizing role is played here by the
vector mesons.

One may decouple the vector mesons and/or dilaton field
from the other by making the corresponding particles infinitely heavy.
In this limit, the dilaton field is frozen to its vacuum
value $\chi=f_\chi$ and rho mesons are constrained to
\begin{equation}
i\vec{\tau}\cdot\vec{\rho}_\mu = \frac{1}{g}
( \partial_\mu \xi_L^{}\xi^\dagger_L
 +\partial_\mu \xi_R^{}\xi^\dagger_R )
\end{equation}
as kind of composite particles made of two pions. Then, the model
Lagrangian (\ref{L:pcrw}) reduces to (\ref{L:pc}) or (\ref{L:p}),
where the kinetic energy term of the rho vectors becomes the
Skyrme term.

\section{Dense Skyrmion Matter}
These {\em nonlinear} meson Lagrangian supports soliton solutions,
skyrmions, carrying nontrivial topological winding numbers.
Once we accept Skyrme's conjecture
of interpreting the winding number as the baryon number,
we may describe a dense baryonic matter as a system made of
many skyrmions.

In the {\em classical} picture, the lowest energy state of the
multi-skyrmion system is a crystal and there has been intensive
work in late 80's with the model containing pions only
\cite{SkyrmionCrystal}. Well-separated two skyrmions have lowest
energy when they are relatively rotated in the isospin space about
an axis perpendicular to the line joining their centers. We can
imagine that the lowest energy state of skyrmion matter in a
relatively low density is in an FCC(face centered cubic) crystal
phase where well localized single skyrmions are arranged on each
lattice cite in a way that 12 nearest skyrmions have such lowest
energy relative orientations. At higher density, on the other
hand, skyrmion tails start overlapping and the system undergoes a
phase transition to a more symmetric configuration the so called
the ``half-skyrmion" cubic crystal. There, one half of the baryon
number carried by the single skyrmion is concentrated at original
FCC site where $U=-1$ while the other is concentrated on the links
where $U=+1$. Now, the system has an additional symmetry with
respect to $U \rightarrow -U$, which results in the vanishing of
$\langle\sigma\rangle(U\equiv\sigma+i\vec{\tau}\cdot\vec{\phi})$.
In the literature, it is often interpreted as the restoration of
the chiral symmetry. However, it is only the average value of
$\sigma$ over the space that vanishes, while the chiral circle
still has a fixed radius $f_\pi$. We call this phase as
``pseudogap" phase to distinguish it from the genuine chiral
symmetry restored phase, where the chiral circle shrinks to a
point at $\sigma=0,\vec{\phi}=0$.

In the model with dilaton field which plays more or less a role of
the ``radial" field for $U$, the restriction on chiral radius
becomes released. The pseudogap phase still remains as a transient
process unless the dilaton mass is sufficiently
small\cite{LPRV03b}. Shown in Fig.1 is a typical numerical results
on the average values of $\sigma$ and $\chi/f_\chi$. In the
figure, the explicit scale of the baryon number density, which
depends strongly on the model parameters, is not shown. One can
see that, as the density increases, the average value of $\sigma$
drops quickly and reaches zero at the density $\rho_p$, where the
system changes to the pseudogap phase. As we increase the density
further, the system remains in the pseudogap phase for a while but
the average value of $\chi/f_\chi$ keeps going down slowly. At the
density $\rho_c$, the $\langle \chi/f_\chi\rangle \neq 0$ phase
and the $\langle \chi/f_\chi\rangle = 0$ phase have the same
energy. Then, at higher density than $\rho_c$, the latter comes to
have lower energy and finally chiral symmetry is restored.

\begin{figure}
\begin{center}
\includegraphics[width=11pc,angle=270]{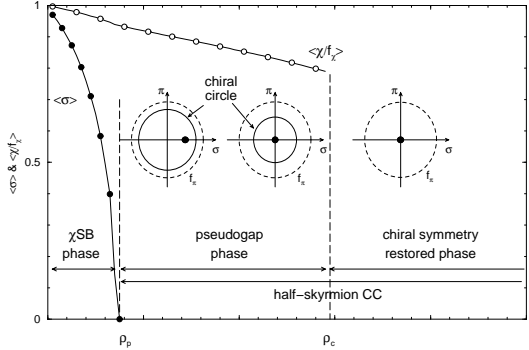}
\end{center}
\vskip -4ex
\caption{Average values of $\sigma$ and $\chi/f_\chi$
of the lowest energy crystal configuration at a given baryon number
density.}
\end{figure}

The vector mesons (especially the omega meson) also play important
roles in dense skyrmion matter. Numerical results obtained in
various models are presented in Figure 2. In the $\pi\rho$ model,
as the density of the system increases (and the lattice constant
decreases), the system energy per baryon, $E/B$, changes slightly.
Its value is close to the energy of a single skyrmion up to
somewhat high densities. The main reason for this is that the size
of the skyrmion is very small because of the absence of strong
repulsive terms in the model. Thus, the skyrmions in the lattice
interact only at very high densities where their tails overlap. In
the $\pi\rho\chi$ model without $\omega$, the dilaton field plays
an important role. Skyrmion matter undergoes an abrupt phase
transition at the density where the expectation value of the
dilaton field vanishes $\langle \chi \rangle=0$.

In the  $\pi \rho \omega \chi$ model, the situation changes
dramatically. Above all, $\omega$ provides a strong repulsion
which inflates each single skyrmion. The tails of the bigger
skyrmions overlap providing attraction to the system in the
intermediate range.

In both the $\pi\rho\omega$ and the $\pi\rho\omega\chi$ models, at
high density, the interaction reduces $E/B$ to 85\% of the $B=1$
skyrmion mass. This value should be compared with 94\% in the
$\pi\rho$ model. In the $\pi\rho\chi$-model, $E/B$ goes down to
74\% of the $B=1$ skyrmion mass, but in this case it is due to the
dramatic behavior of the dilaton field. On the other hand, the
omega suppresses the role of the dilaton field. It could provide
only a small attraction at intermediate densities. Moreover, the
phase transition towards its vanishing expectation value, $\langle
\chi\rangle=0$ does not take place. Instead, its value grows at
high density!

The reason for this can be found in the role played by omega in
(\ref{L:pcrw}). In the static configuration, omega produces a
potential, whose source is the baryon number density, which
mediates the self-interaction energy of the baryon number
distribution. Thus, unless it is screened properly by the omega
mass, the periodic source filling infinite space will lead to an
infinite self-energy. To reduce the energy of the system, the
effective $\omega$ mass must grow at high density, for which
$\chi$ must grow too. Note the factor $(\chi/f_\chi)^2$ in the
omega mass term in Lagrangian (\ref{L:pcrw}).

\begin{figure}
\begin{center}
\includegraphics[width=13pc,angle=270]{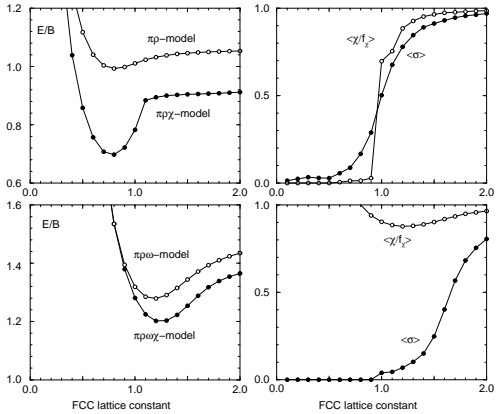}
\end{center}
\vskip -4ex
\caption{Numerical results in the model with vector mesons
and dilaton field.}
\end{figure}

On the other hand, these classical crystalline structures are
quite far from normal nuclear matter which is known to be a Fermi
liquid at low temperature. In order for skyrmion matter to be
identified with nuclear matter we have to quantize and thermalize
the classical system. Since it is a system of solitons in a meson
field theory, it is not sufficient to quantize the meson
fluctuations. We need to introduce and quantize proper collective
variables not only to complement the broken symmetries of the
whole skyrmion system but also to describe the dynamics of the
single skyrmions in order to obtain a realistic picture of nuclear
matter. For extended objects, we may need an infinitely large
number of dynamical variables such as the positions of their
center of mass, their relative orientations, their sizes, their
deformations, etc. Among them, those degrees of freedom that
describe translations and rotations of the single skyrmion play
the dominant role at low energy. Thus, we need at least 6
variables for each skyrmion. For a multi-skyrmion system, the
simplest way of introducing collective variables for the position
and orientation of each single skyrmion is through the use of the
product Ansatz, the old idea of Skyrme\cite{Sk62}. There, a
multi-skyrmion solution can be approximated by products of single
skyrmion solutions centered at the corresponding positions and
rotated to have the corresponding orientation. However, the
product Ansatz works well only when the skyrmions are sufficiently
separated. Furthermore, due to the non-commutativity of the matrix
products, it is difficult to use it in multi-skyrmion systems.

   Another scheme which can be used to study multi-skyrmion systems
is the Antiyah-Manton Ansatz \cite{AM89}. In this scheme,
skyrmions of baryon number $N$ are obtained by calculating the
holonomy of Yang-Mills instantons of charge $N$, which has been
used in describing successfully few-nucleon systems and also
nuclear matter. One advantage of the Atiyah-Manton ansatz is that
it provides a natural framework to introduce the proper dynamical
variables for the skyrmions through the parameters describing the
multi-instanton configuration. Contrary to the non-commutative
product ansatz, some multi-instanton solutions are given in a
commutative manner. Furthermore, multi-instanton solutions have
been investigated widely and many useful solutions have been
found.

We have tried this idea in Ref.\cite{PMRV02}. 'tHooft ansatz on
multi-instanton solution is adopted to produce the FCC {\em
instanton crystal}. To avoid the divergence coming from the
infinite number of instantons and to incorporate the relative
orientations to each instantons, the ansatz is slightly modified.
Then, the Antiyah-Manton procedure is applied to transform the
instanton crystal into a skyrme crystal. As expected, the
resulting skyrmion crystal is FCC for low baryon number density.
At high density, however, it becomes a {\em half-skyrmion CC}
(approximately).

One great advantage of this procedure to generate the skyrmion
crystal is that it is really made up of single objects located at
specified positions and with specified rotations. Now, we can, for
example, move a single skyrmion and investigate how the system
changes. Shown in Fig.3 is the potential energy $V(d)$ when a
single skyrmion is moved away by a distance $d$ from its stable
position. Two extreme cases are shown. In the case of a dense
system ($L_{\mbox{FCC}}=5.0$), the energy changes abruptly. For
small $d$, it is almost quadratic in $d$. It implies that the
dense system is in the crystal phase. On the other hand, in the
case of dilute system ($L_{\mbox{FCC}}=10.0$), the system energy
remains almost constant up to some distance $d$, which implies
that the system is in a gas or liquid phase. If we allow all the
variables to vary freely to seek the minimum energy configuration,
the system will end up in a disordered phase, in which a few
skyrmions will form a finite cluster. Furthermore, we will be able
to investigate the thermal properties of the skyrmion system and
quantize those collective variables.

\begin{figure}
\begin{center}
\includegraphics[width=11pc,angle=270]{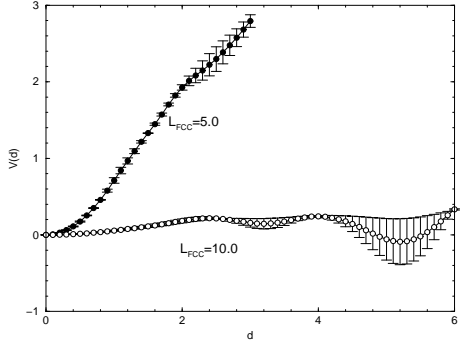}
\end{center}
\vskip -4ex
\caption{The energy cost $V(d)$ to shift a single skyrmion from
its stable position by an amount $d$ along the edge of FCC lattice.}
\end{figure}

\section{Mesons in Dense Matter}

By using the same model Lagrangian defined at zero baryon number
density, we may investigate the properties of the mesons in dense
matter. The Lagrangians (\ref{L:p})-(\ref{L:pcrw}) are originally
constructed to describe the meson dynamics in vacuum, which is
defined as
\begin{equation}
U=1, \hskip 1em \chi=f_\chi^{} \hskip 1em \mbox{and} \hskip 1em
\vec{\rho}_\mu^{}=\omega_\mu^{}=0.
\end{equation}
The fluctuations on top of this vacuum can describe the
corresponding particles in zero baryon number space, for which we
determine the physical parameters of the model Lagrangian, such as
the particle masses, decay constants and coupling constants.

Now, as we have described in the previous section, we have another
classical background configuration for the matter with non
vanishing baryon number density. Let's denote the solution as
$U_{(0)}^{}$, $\chi_{(0)}^{}$, $\vec{\rho}^\mu_{(0)}$ and
$\omega^\mu_{(0)}$, respectively. We can incorporate the meson
fluctuation on top of this background as
\begin{equation}
\begin{array}{l}
U=\sqrt{U_{\pi}^*} U_{(0)} \sqrt{U_{\pi}^*}, \hskip 2em
\chi=\chi_{(0)}+\chi^*, \\
\vec{\rho}^\mu =\vec{\rho}^\mu_{(0)} + \vec{\rho}^{*\mu}, \hskip 2em
\omega^\mu=\omega^\mu_{(0)}+\omega^{*\mu},
\end{array}
\label{fluct}\end{equation}
where the `starred' fields are describing the corresponding particles
{\em in medium.}

We will illustrate the basic strategy of the approach by using the
simplest model with only pions. Substituting eqs.(\ref{fluct})
into the Lagrangian density (\ref{L:p}) and keeping the terms up
to the second order in the fluctuating fields, we obtain
\begin{equation}
\begin{array}{rcl}
{\cal L}^*
&=&
\frac12 G_{ab}(\vec{r}) \partial_\mu \pi^{*}_a \partial^\mu \pi^{*}_b \\
&&
+ \frac12 m_\pi^2 \sigma(\vec{r}) \pi^{*}_a \pi^{*}_a+\cdots,
\end{array}
\label{L:p*}\end{equation} where $G_{ab}(\vec{r})$ and
$\sigma(\vec{r})$ are the background potentials provided by the
dense skyrmion matter. For the fluctuating pions in free space,
both potentials are just 1. This Lagrangian tells us how the
properties of the pions change in the dense matter. For example,
$G_{ab}(\vec{r})$ in front of the pion kinetic term can be
absorbed into the wavefunction renormalization, which can be
interpreted as a ratio between the {\em local effective} pion
decay constant in dense medium to that in free space. The pion
mass term gets a similar correction from dense matter. To the
first order in the background potentials, we can translate
(\ref{L:p*}) into an effective Lagrangian for the pions in dense
medium with {\em in-medium} physical parameters as
\begin{equation}
\frac{f^*_\pi}{f_\pi}
 = \sqrt{\langle G_{aa} \rangle},
\hskip 2em
\frac{m^*_\pi}{m_\pi}
 = \sqrt{\frac{\langle \sigma\rangle}{\langle G_{aa} \rangle}}.
\end{equation}

In Figure 4 we show the ratios of the in-medium parameters of
pions relative to its free-space values. The pion mass is almost
constant at low density but decreases as the density increases to
much higher values. The pion decay constant drops quite fast but
after some density it stays with value  $\sim 2/3$ indicating that
the system is in the pseudogap phase mentioned in Section 3. A
similar process can be applied after incorporating the
dilaton\cite{LPRV03a}, where the vanishing of $\chi$ can restore
the chiral symmetry completely.

\begin{figure}
\begin{center}
\includegraphics[width=11pc,angle=270]{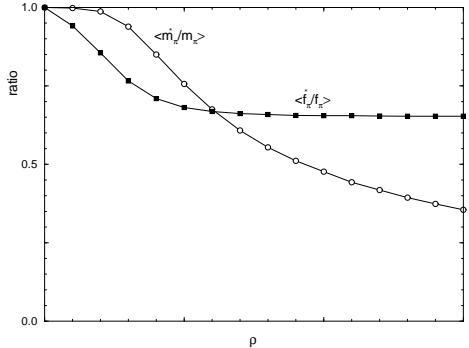}
\end{center}
\vskip -4ex
\caption{In medium properties of the pions
as a function of the baryon number density. }
\end{figure}

One may treat the interaction of the fluctuating fields with the
background potentials in a more systematic way. As an example, in
Ref.\cite{LPRV03b} the in-medium pion velocity is studied, where
the background interactions are taken into account up to the
second order. The breakdown of the Lorentz symmetry due to the
presence of medium makes the pion velocity deviate from that in
free space.

\section{Summary}
We have developed a unified approach to dense matter within the
Skyrme philosophy, where systems of baryons and mesons can be
described by a single Lagrangian. In our approach dense baryonic
matter is approximated by skyrmion matter in the lowest energy
configuration for a given baryon number density. By incorporating
in it fluctuating mesons we can get some insight on meson dynamics
in a dense medium. Our approach enables us to study this dynamics
beyond the first order in the baryon number density. One can
continue to work in this direction by incorporating more degrees
of freedom, by improving the way of treating matter beyond the
crystal solution, and so on.

However, before closing the presentation, we must clearly lay down
the scope of our work. We do not claim that the results obtained
at present describe reality. The most fundamental problem we phase
is that our ``ground state" for matter is a crystal not a Fermi
liquid. Our aim has been to assume a state for matter, given by a
classical solution of a theory considered to be valid at large
$N_c$, and have studied the implications for its excitations. Our
work should be taken as representing the first step towards a more
realistic treatment of a dense matter theory.

\end{document}